\documentclass[preprint]{aastex631}
\usepackage{hyperref}
\newcommand{\msun}{M$_\odot$}

\begin{document}

\title{A speckle search for the outer companion of KIC 9832227}

\author[0000-0002-1206-1930]{Ricardo Salinas}
\affiliation{Departamento de Astronom\'ia, Universidad de La Serena, La Serena, Chile}
\author[0000-0002-2532-2853]{Steve B. Howell}
\affiliation{NASA Ames Research Center, Moffett Field, CA 94035, USA}

\begin{abstract}
We present Gemini-N/'Alopeke speckle observations of KIC 9832227, a system originally predicted to become a red nova. The diffraction limited observations do not find an outer companion between 11 and 678 AU that could be responsible for the period changes of the system.
\end{abstract}

\keywords{Contact binary stars(297)	--- Speckle interferometry(1552)}

\section{Introduction} \label{sec:intro}

KIC 9832227 is a contact binary with a changing orbital period which was originally thought to be exponentially declining, predicting a merger as a red nova by 2022 \citep{molnar17}. This prediction was later negated based on a data re-analysis \citep{socia18}, but the origin of the change in the period has remained unknown, with an outer companion on a hierarchical triplet being the most likely culprit \citep{kovacs19}. Period changes in contact binaries, both positive and negative, are not unheard of. One illustrative example is the famous contact binary star W UMa whose changing period was well documented in the 1970's \citep[e.g.][]{cester72,cester76}.

In this \textit{Research Note} we exploit diffraction limit observations with the Gemini North 8.1m telescope to probe the existence of outer companions to KIC 9832227.

\section{Observations} \label{sec:observations}

Observations were carried out with the speckle interferometer 'Alopeke \citep{scott21}, mounted at the Gemini North Telescope in MaunaKea. 20$\times$1000$\times$60ms exposures in medium band filters centered at 562nm and 832nm were obtained simultaneously under Gemini program GN-2022A-LP-205 on May 10, 2022. Data reduction of the speckle images is explained in \citet{howell11}.

\section{Results and analysis} \label{sec:results}

\begin{figure}
\begin{minipage}[]{0.5\textwidth}
\includegraphics[scale=0.59]{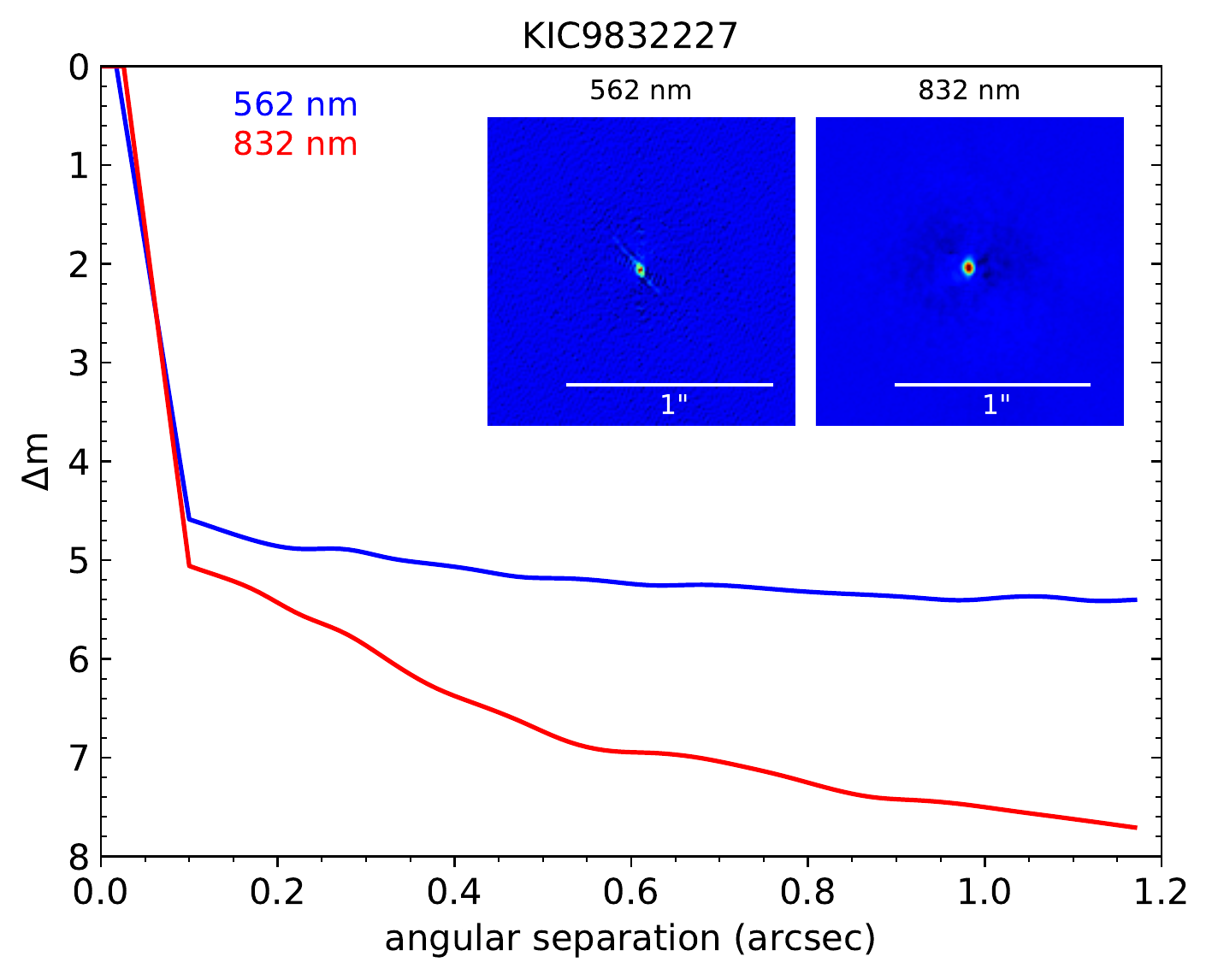}
\end{minipage}
\begin{minipage}[]{0.5\textwidth}
\includegraphics[scale=0.5]{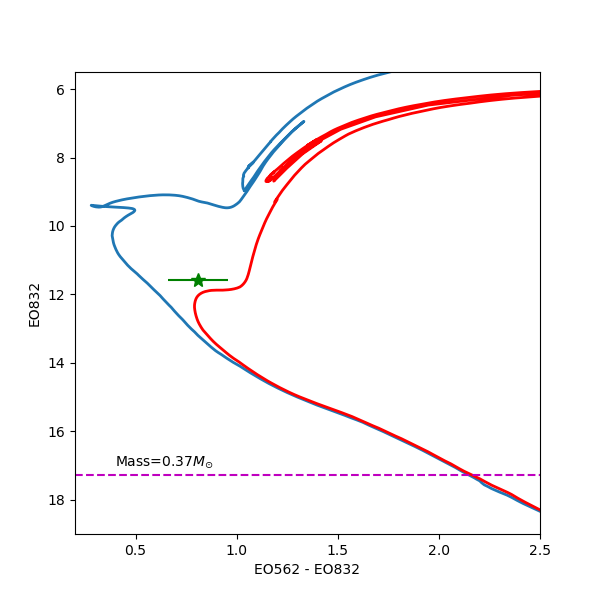}
\end{minipage}
\caption{\textbf{Left panel:} 5-$\sigma$ detection curves for the 'Alopeke observations at 562nm (in blue) and 832nm (in red). The insets show the reconstructed images in both filters. No companions to KIC 9832227 above these limits are found. \textbf{Right panel}: A color magnitude diagram in the 'Alopeke filters for representative 1 Gyr (blue) and 10 Gyr (red) isochrones from MIST \citep{choi16} with [Fe/H]=--0.04 \citep{molnar17}. The position of KIC 9832227 in these filters is shown in green. The dashed magenta line indicates the detection limit at 100mas based on the 832nm detection curve.}
\label{fig:alopeke}
\end{figure}

The speckle observations detected no companions around KIC 9832227 between 20 mas and 1.2 arcsec, above the 5--$\sigma$ contrast curves shown in Fig. \ref{fig:alopeke} (left panel). These angular limits equal a 11.3 AU to 678 AU range when using the distance to the system derived by \citet{molnar17} of 565 pc. \textit{Gaia} DR3 gives a slightly farther distance, \verb+distance_gspphot+=619.92 pc \citep{gaiadr3}, but given the large astrometric noise of the observation we use the \citet{molnar17} distance in the following. The relative magnitude contrast curves were calibrated into absolute magnitudes using the \textit{Gaia} magnitude for the system ($G=12.2$), a G0V spectrum from the \citet{pickles98} library (consistent with the temperature of 5\,828 K measured by \cite{molnar17}) and the 'Alopeke transmission filter curves\footnote{https://www.gemini.edu/instrumentation/alopeke-zorro/components} using \verb+pyphot+\footnote{\href{https://mfouesneau.github.io/pyphot/}{https://mfouesneau.github.io/pyphot}}.
 
To transform the contrast curves into mass limits for any companion we used the MIST stellar models \citep{choi16} transformed into the 'Alopeke filters using SPISEA \citep[][Fig. \ref{fig:alopeke}, right panel]{hosek20}\footnote{\href{https://spisea.readthedocs.io/}{https://spisea.readthedocs.io/}}. At a distance of 100 mas, the 5-$\sigma$ detection limit in the 832nm filter equals 17.27 mag which transforms into a mass of 0.37 \msun; at 1\arcsec\, the detection limit of 20.02 mag yields a mass 0.13 \msun, irrespective of the age of the system and for a metallicity of [Fe/H]=--0.04 \citep{molnar17}.

Even though these observations cannot probe the most up-to-date prediction for the position of the third member of this putative hierarchical system \citep[$P\sim$13.5 years; $\sim8.2$ AU,][]{kovacs19}, they can rule out a possible non-degenerate companion at a larger distance, for example, the component D in the original model of \citet{molnar17}, expected to have a period longer than 25 years.

\begin{acknowledgments}
RS thanks Geza Kovacs for a useful discussion and Camila Beltrand for help with matplotlib.  Observations in this paper made use of the High-Resolution Imaging instrument 'Alopeke. 'Alopeke was funded by the NASA Exoplanet Exploration Program and built at the NASA Ames Research Center by Steve B. Howell, Nic Scott, Elliott P. Horch, and Emmett Quigley.  'Alopeke was mounted on the Gemini North telescope of the international Gemini Observatory, a program of NSF’s OIR Lab, which is managed by the Association of Universities for Research in Astronomy (AURA) under a cooperative agreement with the National Science Foundation on behalf of the Gemini Observatory partnership: the National Science Foundation (United States), National Research Council (Canada), Agencia Nacional de Investigaci\'{o}n y Desarrollo (Chile), Ministerio de Ciencia, Tecnolog\'{i}a e Innovaci\'{o}n (Argentina), Minist\'{e}rio da Ci\^{e}ncia, Tecnologia, Inova\c{c}\~{o}es e Comunica\c{c}\~{o}es (Brazil), and Korea Astronomy and Space Science Institute (Republic of Korea).
\end{acknowledgments}

\vspace{5mm}
\facilities{Gemini:Gillett('Alopeke)}

\software{pyphot, SPISEA \citep{hosek20}, matplotlib \citep{Hunter07}}
          
\bibliography{kic}{}
\bibliographystyle{aasjournal}
\end{document}